# Comparison of different Methods for Univariate Time Series Imputation in R

*by Steffen Moritz, Alexis Sardá, Thomas Bartz-Beielstein, Martin Zaefferer and Jörg Stork*

**Abstract** Missing values in datasets are a well-known problem and there are quite a lot of R packages offering imputation functions. But while imputation in general is well covered within R, it is hard to find functions for imputation of univariate time series. The problem is, most standard imputation techniques can not be applied directly. Most algorithms rely on inter-attribute correlations, while univariate time series imputation needs to employ time dependencies. This paper provides an overview of univariate time series imputation in general and an in-detail insight into the respective implementations within R packages. Furthermore, we experimentally compare the R functions on different time series using four different ratios of missing data. Our results show that either an interpolation with seasonal kalman filter from the *zoo* package or a linear interpolation on seasonal loess decomposed data from the *forecast* package were the most effective methods for dealing with missing data in most of the scenarios assessed in this paper.

## Introduction

Time series data can be found in nearly every domain, for example, biology (Bar-Joseph et al., 2003), finance (Taylor, 2007), social science (Gottman, 1981), energy industry (Billinton et al., 1996) and climate observation (Ghil and Vautard, 1991). But nearly everywhere, where data is measured and recorded, issues with missing values occur. Various reasons lead to missing values: values may not be measured, values may be measured but get lost or values may be measured but are considered unusable. Possible real life examples are: markets are closed for one day, communication errors occur or a sensor has a malfunction. Missing values can lead to problems, because often further data processing and analysis steps rely on complete datasets. Therefore missing values need to be replaced with reasonable values. In statistics this process is called imputation.

Imputation is a huge area, where lots of research has already been done. Examples of popular techniques are Multiple Imputation (Rubin, 1987), Expectation-Maximization (Dempster et al., 1977), Nearest Neighbor (Vacek and Ashikaga, 1980) and Hot Deck (Ford, 1983) methods. In the research field of imputation, univariate time series are a special challenge. Most of the sufficiently well performing standard algorithms rely on inter-attribute correlations to estimate values for the missing data. In the univariate case no additional attributes can be employed directly. Effective univariate algorithms instead need to make use of the time series characteristics. That is why it is senseful to treat univariate time series differently and to use imputation algorithms especially tailored to their characteristics.

Until now only a limited number of studies have taken a closer look at the special case of univariate time series imputation. Good overview articles comparing different algorithms are yet missing. With this paper we want to improve this situation and give an overview about univariate time series imputation. Furthermore, we want to give practical hints and examples on how univariate time series imputation can be done within R [1].

The paper is structured as follows: Section Univariate Time Series defines basic terms and introduces the datasets used in the experiments. Afterwards section Missing Data describes the different missing data mechanisms and how we simulate missing data in our experiments. Section Univariate time series imputation gives a literature overview and provides further details about the R implementations tested in our experiments. The succeeding section explains the Experiments in detail and discusses the results. The paper ends with a short Summary of the gained insights.

## Univariate Time Series

### Definition

A *univariate time series* is a sequence of single observations $o_1, o_2, o_3, ... o_n$ at successive points $t_1, t_2, t_3, ... t_n$ in time. Although a univariate time series is usually considered as one column of observations, time is in fact an implicit variable. This paper only concerns equi-spaced time series. Equi-spaced means, that time increments between successive data points are equal $|t_1 - t_2| = |t_2 - t_3| = ... = |t_{n-1} - t_n|$.

---

[1] The R code we used is available online in the GitHub repository http://github.com/SpotSeven/uniTSimputation.

For representing univariate time series, we use the *ts {stats}* time series objects from base R. There are also other time series representation objects available in the packages **xts** (Ryan and Ulrich, 2014), **zoo** (Zeileis and Grothendieck, 2005) or **timeSeries** (Team et al., 2015). While *ts* objects are limited to regularly spaced time series using numeric time stamps, these objects offer additional features like handling irregular spaced time series or POSIXct timestamps. Since we do not need these additions, we chose to use *ts* objects for our experiments. Also important to note is, that we assumed that the frequency (number of observations per season) of the time series is either known or set to one.

```
#' Example for creating a ts object with a given  frequency
#' Working hours of an employee recorded Monday till Friday
  workingHours <- c(8.2, 8.2, 7.9, 8.3, 7.2, 8.2, 8.5, 7.2, 8.7,  7.1)
  tsWorkingHours <- ts(workingHours, frequency = 5)
```

### Data Characteristics

For the later experimental part of the paper, we decided to compare performance of the imputation algorithms on four different time series datasets provided in the **TSA** (Chan and Ripley, 2012) package. One reason for choosing these datasets (besides other reasons explained in the following paragraphs) is, that these are well-known and frequently used in literature.

The datasets we chose are (see also figure 1):

- **airpass** - Monthly total international airline passengers from 01/1960 - 12/1971
- **beersales** - Monthly beer sales in millions of barrels, 01/1975 - 12/1990
- **SP** - Quarterly S&P Composite Index, 1936Q1 - 1977Q4
- **google** - Daily returns of the google stock from 08/20/04 - 09/13/06

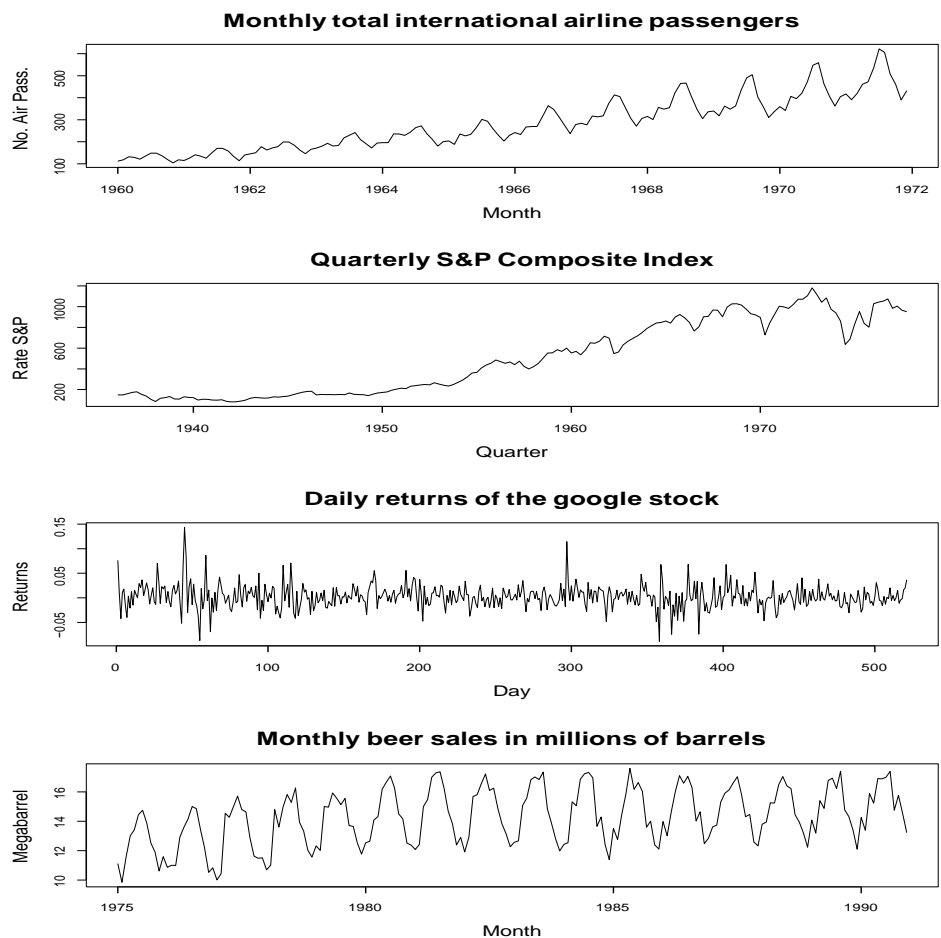

**Figure 1:** Time series datasets used for the experimental part of the paper.

```
#' Code for making the datasets available and plotting  them
  library("TSA")
  data(list = c("airpass", "SP", "google", "beersales"))
  par(mfrow=c(4,1), cex.main = 1.8, cex.lab = 1.4)
  plot(airpass,main = "Monthly total international airline passengers", xlab="Month",
      ylab = "No. Air Pass.")
  plot(SP,main ="Quarterly S&P Composite Index",  xlab="Quarter",
      ylab = "Rate S&P")
  plot(google,main = "Daily returns of the google stock", xlab="Day",
      ylab = "Returns")
  plot(beersales, main = "Monthly beer sales in millions of barrels", xlab="Month",
      ylab ="Megabarrel")
```

Choosing the number of considered datasets is a trade-off between creating meaningful results and keeping complexity low. Usually, results are more significant the more different datasets are compared. On the other hand, it is not possible to make a very detailed analysis for a higher number of datasets. Instead of comparing on a high number of datasets we chose four datasets, which we thought are representative for different time series characteristics. In the following the different characteristics of the time series are explained.

Two common approaches to describe and examine time series are *autocorrelation* analysis and separation into *trend*, *seasonal* and *irregular* components.

**Decomposition**
Time series data can show a huge variety of patterns and, for analysis, it can be useful to isolate these patterns in separate series. Time series decomposition seeks to split the time series into single component series each representing a certain characteristic or pattern. (the original time series can later on be reconstructed by additions or multiplications of these components)

There are typically three components of interest:

- **trend** component - expresses the long term progression of the time series (secular variation). This means there is a long term increase or decrease in the mean level. The trend does not necessarily have to be linear.

- **seasonal** component - reflects level shifts that repeat systematically within the same period (seasonal variation). This means there is a common pattern repeating for example every month, quarter of the year or day of the week. Seasonality is always of a fixed and known period.

- **irregular** component - describes irregular influences. These are the residuals, after the other components have been removed from the time series. The irregular component may be, but is not necessarily completely random. It can show autocorrelation and cycles of unpredictable duration.

There are different techniques for doing the decomposition into components. The basic method is a decomposition by moving averages, but there exist more robust and versatile methods like STL (Seasonal and Trend decomposition using Loess) decomposition (Cleveland et al., 1990). In figure 2 a STL decomposition of the air passengers dataset can be seen.

Considering trend and seasonal influences is very important for the results of time series analysis. Getting trend and seasonal effects right can improve forecasting and even imputation results a lot. Thus decomposition is a popular technique and especially often used for seasonal adjustment.

Looking again at figure 1, it can be seen why we chose exactly these four datasets. The google dataset is showing no trend and no seasonality at all, appearing to be nearly random. A quite typical behavior for short time observations for financial time series. The SP dataset is showing a clear trend but no seasonality. This is a really common long term observation for stock prices. In the beer sales series it can be seen that there is more beer consumption in the hot summer months. But the overall beer consumption over the years stays nearly constant. Meanwhile the amount of air passengers increase significantly over the years, but also shows seasonal shifts.

Hence, the chosen data sets have the following features:

- no trend, no seasonality (google)
- trend, no seasonality (SP)
- no trend, seasonality (beersales)
- trend, seasonality (airpassengers)

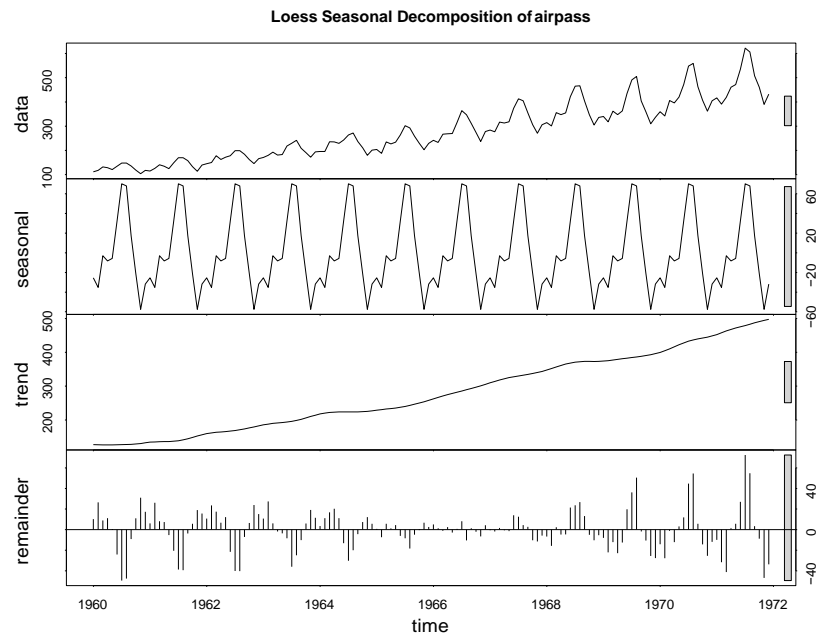

**Figure 2:** Loess seasonal decomposition of airpass time series dataset.

```
#' Code for doing a seasonal loess decomposition
  stlDecomp <- stl(airpass, s.window = "periodic")
  plot(stlDecomp)

#' Code for doing a seasonal Moving Averages  decomposition
  library(forecast)
  maDecomp <- decompose(airpass)
  plot(maDecomp)
```

**Autocorrelation**
Autocorrelation, also called serial correlation, is a measure of the internal correlation within a time series. It is a representation of the degree of similarity between the time series and a lagged version of itself. The calculation process is similar to getting correlations between two different time series, except that one time series is used twice, once in its original form and once in a lagged version.

The original idea of measuring autocorrelation is that forecasting (and also imputation) of a time series is possible because future usually depends on the past. High autocorrelation values mean that the future is strongly correlated to the past. Thus autocorrelation can be an indicator for the ability to create reliable forecasts and imputations.

When the autocorrelation is computed the resulting numbers range from +1 to -1. A value of one means that there is a perfect positive association, a value of minus one means that there is a perfect negative association and zero means there is no association.

In figure 3 autocorrelation plots of the lags for the time series used in our experiments can be seen. The blue striped line marks the limit, under which the autocorrelation is not statistically significant. Looking at the air passengers and the SP series, it can be seen that there is a strong positive autocorrelation for both of them. Beersales shows repeating patterns of positive and negative autocorrelations, typical for seasonality. Remarkable is the google dataset, which by contrast shows no autocorrelation over the significance line. This dataset was chosen to see how imputation algorithms will perform for series that have no clear patterns and show white noise characteristics.

```
#' Code for creating autocorrelation plot
#' No plot function call, because it is already called within  acf
  par(mfrow=c(2,2), cex.main = 1.8, cex.lab =  1.4)
  acf(airpass)
  acf(SP)
  acf(google)
  acf(beersales)
```

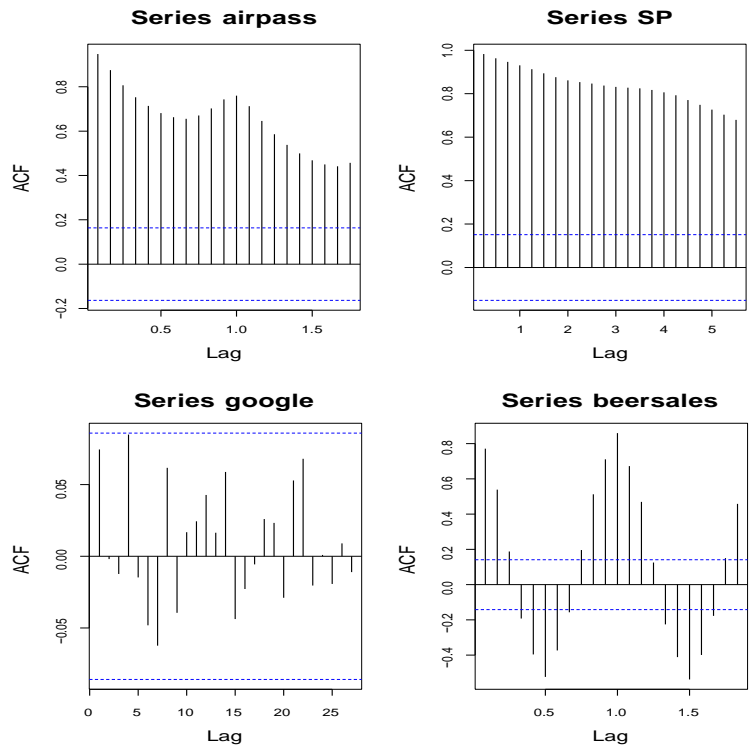

**Figure 3:** Autocorrelation plot for the time series used in the experiments.

## Missing Data

### Missing Data Mechanisms

Depending on what causes missing data, the gaps will have a certain distribution. Understanding this distribution may be helpful in two ways. First, it may be employed as background knowledge for selecting an appropriate imputation algorithm. Second, this knowledge may help to design a reasonable simulator, that removes missing data from a test set. Such a simulator will help to generate data where the true values (i.e., the potentially ideal imputation data) is known. Hence, the quality of an imputation algorithm can be tested.

Missing data mechanisms can be divided into three categories 'Missing completely at random' (*MCAR*), 'Missing at random' (*MAR*) and 'Not missing at random' (*NMAR*). In practice, assigning data-gaps to a category can be blurry, because the underlying mechanisms are simply unknown. While MAR and MNAR diagnosis needs manual analysis of the patterns in the data and application of domain knowledge, MCAR can be tested for with t-test or Little's test (Little, 1988). In the R package **MissMech** (Jamshidian et al., 2014) a function for MCAR diagnosis can be found.

The vast majority of missing data methods require MAR or MCAR, since the missing data mechanism is said to be *ignorable* for them (Rubin, 1976). Since MAR enables imputation algorithms to employ correlations with other variables, algorithms can achieve better results than for MCAR. NMAR is called non-ignorable, because in order to do the imputation a special model for why data is missing and what the likely values are has to be included.

For univariate time series the picture for missing data mechanisms looks slightly different. At first view there is only one variable in the data. But indeed time (which is implicitly given) has to be treated like a variable when determining the missing data mechanism of a dataset. Another difference is that time series imputation algorithms do not need to rely solely on covariates for missing value estimation, they can also use time series characteristics. This makes estimating missing values for MCAR data a lot of easier. For univariate time series imputation MAR and MCAR is nearly the same.

### Missing completely at random (MCAR)

In MCAR there is no systematic mechanism on the way the data is missing. Missing data points occur entirely at random. This means there are two requirements: First, the probability for a certain observation being missing is independent from the values of other variables. Second, the probability for an observation being missing is also independent of the value of the observation itself. Since there are no other variables existent for univariate time series (except time as implicit variable), condition one can be simplified to: The probability for a certain observation being missing is independent of the point of time of this observation in the series.

**Example**: Sensor data is recorded 24/7 from a field test and sent via radio signals to a back-end. Due to unknown reasons and on random occasions sometimes the transmission fails.

$$P(r|Y_{observed}, Y_{missing}) = P(r)$$

**Missing at random (MAR)**

Like in MCAR, in MAR the probability for an observation being missing is also independent of the value of the observation itself. But it is dependent on other variables. Since there are no other variables other than time (implicitly given) for univariate time series, it can be said, that in MAR the probability for an observation being missing is dependent of the point in time of this observation in the series.

**Example**: Machine sensor data are more likely to be missing on weekends (since it is shut down on some weekends).

$$P(r|Y_{observed}, Y_{missing}) = P(r|Y_{observed})$$

**Not missing at random (NMAR)**

NMAR observations are not missing in a random manner. The data are not MCAR and not MAR. That means, the probability for a observation being missing depends on the value of the observation itself. Furthermore the probability can (but must not necessarily) be dependent on other variables (point of time for univariate time series).

**Example**: Temperature sensor gives no values for temperatures over 100 degrees.

$$P(r|Y_{observed}, Y_{missing}) = P(r|Y_{observed}, Y_{missing})$$

**Simulation of Missing Data**

Evaluating performance of imputation algorithms has one difficulty: Comparing the results on real missing data is not possible, since the actual values are in fact missing. Thus it will never emerge how much the imputed vales differ from the real values. A performance comparison can only be done for simulated missing data. This means a complete series is taken and data points are artificially removed. Later on, imputed values and real values can be compared. The characteristics of the function created for the experiments are shown in table 1.

| Missing Data Mechanism | Missing Data Distribution | Amount of Missing Data |
|---|---|---|
| *MCAR* | *exponential* | *adjustable* |

**Table 1:** Characteristics Missing Data Simulation

These characteristics were not chosen arbitrarily. We decided to use six hundred real life univariate time series as archetype for the function. These time series datasets provided by an industrial partner are recordings of sensor values from field tests. The missing observations in these time series result from unspecified transmission problems. The exponential distribution we discovered in this datasets is very common for a lot of real life applications. For example it is used for modeling time distance between incoming phone calls in call centers (Brown et al., 2005) and also reliability engineering makes extensive use of it (Marshall and Olkin, 1967).

```
#' Code of the missing data simulation function
#' @param data - univariate time series
#' @param rate - lambda value for exponential distribution
```

```
#' @param seed - random seed used for the function

  create.missing <- function(data, rate, seed=NULL) {

      ## Only for time series
      if (class(data) != "ts") {
          stop("Provided data is not a time series.")
      }

      ## No missing data (pass-through)
      if (rate == 0) {
          return(data)
      }

      ## Save original parameters
      t <- time(data)
      f <- frequency(data)

      ##Setting of random seed
      if (!is.null(seed))
          set.seed(seed)

      ## Initialize index
      a <- 0

      ## Indices of removed entries
      tempDelete <- numeric(0)

      while (a < length(data)) {

          ## 'ceiling' is to avoid possible zeros
          a <- ceiling(a + rexp(1, rate))

          if ( a <= length(data) ) {
              data[a] <- NA
              tempDelete <- c(tempDelete, a)
          }
      }

      return(list(data=data, na.ind=tempDelete))
  }
```

## Univariate time series imputation

### Overview Algorithms

As already mentioned in the Univariate Time Series and Missing Data sections, univariate time series are a special imputation case. Instead of covariates like in multivariate datasets, time dependencies have to be employed to perform an effective imputation. Literature about time series or longitudinal data focuses nearly solely on multivariate datasets. Most papers thereby compare algorithms just on one specific dataset, like for example (Honaker and King, 2010) for a political science cross sectional dataset. Especially for longitudinal clinical datasets, imputation algorithm comparisons can be found (Engels and Diehr, 2003), (Spratt et al., 2010), (Twisk and de Vente, 2002), (Spratt et al., 2010). But none consider univariate data. Other articles like (Junninen et al., 2004) consider some univariate algorithms, but do not consider the time series aspects. These simple algorithms like mean imputation usually do not perform well. All in all there is no overview article or overall comparison solely devoted to univariate time series imputation. But there are papers covering some aspects. For example (Kihoro et al., 2013) evaluates the use of ARIMA and SARIMA models for imputation of univariate time series.

Techniques capable of doing imputation for univariate time series can be roughly divided into three categories:

1. **Univariate algorithms**
   These algorithms work with univariate inputs, but typically do not employ the time series character of the dataset. Examples are: mean, mode, median, random sample

2. **Univariate time series algorithms**
   These algorithms are also able to work with univariate inputs, but make use of the time series characteristics. Examples of simple algorithms of this category are locf (last observation carried forward), nocb (next observation carried backward), arithmetic smoothing and linear interpolation. The more advanced algorithms are based on structural time series models and can handle seasonality.

3. **Multivariate algorithms on lagged data**
   Usually, multivariate algorithms can not be applied on univariate data. But since time is an implicit variable for time series, it is possible to add time information as covariates in order to make it possible to apply multivariate imputation algorithms. This process is all about making the time information available for multivariate algorithms. The usual way to do this is via lags and leads. Lags are variables that take the value of another variable in the previous time period, whereas leads take the value of another variable in the next time period.

**Univariate time series imputation in R**

Basically, there are lots of R packages, offering a broad range of imputation tools and functions. There are packages for a lot of different imputation techniques, e.g., imputation based on random forests **missForest** (Stekhoven and Bühlmann, 2012), maximum likehood estimation **mvnmle** (Gross and with help from Douglas Bates, 2012), expectation maximization **mtsdi** (Junger and de Leon, 2012), nearest neighbor observation **yaImpute** (Crookston and Finley, 2007), predictive mean matching **BaBooN** (Meinfelder, 2011), conditional copula specifications **CoImp** (Lascio and Giannerini, 2014) and several other methods. Especially popular are multiple imputation (Rubin and Schenker, 1986) implementations like **mice** (van Buuren and Groothuis-Oudshoorn, 2011), **Hmisc** (Jr et al., 2015) and **Amelia** (Honaker et al., 2011). Many packages like for example **imputeR** (Feng et al., 2014), **VIM** (Templ et al., 2013) and others, offer also imputation frameworks containing several algorithms and tools. There are plenty other R packages not mentioned yet, providing either tools, visualisations or algorithms for imputation (e.g. **mitools**, **HotDeckImputation**, **hot.deck**, **miceadds**, **mi**, **missMDA**, **ForImp** and others). Several other packages also contain imputation functions as side product.

Unfortunately there is no package that is dedicated solely to univariate time series imputation. Evaluating possibilities within R we have taken a look at the following solutions:

**Amelia and mtsdi**

The **Amelia** and **mvnmle** package offer some extra options for time series support. Amelia has two ways of time series support included: One approach works by directly adding lags or leads as covariates, the other one is adding covariates that correspond to time and its polynomials. This works perfectly fine for multivariate datasets, but for univariate inputs the imputation function is not applicable. We tested these packages because the manuals mentioned time series imputation and did not explicitly exclude univariate inputs. But actually, our tests showed that both packages are designed only for processing multivariate data and are not applicable for univariate data. Providing univariate data ends up in an error signaling wrong input.

```
#' Working code for a multivariate dataset
  require("Amelia")
  #load multivariate dataset with NAs (dataset included in Amelia package)
  data(freetrade)

#Impute with polytime option
  out1 <- amelia(freetrade, ts = "year", cs = "country", polytime = 2)

#Impute with lags and leads option
  out2 <- amelia(freetrade, ts = "year", cs = "country", lags = "tariff", leads = "tariff")
```

Error message from **Amelia** amelia function for univariate input:

```
Amelia Error Code: 42
There is only 1 column of data. Cannot impute.
```

The **mtsdi** package provides an EM algorithm based method for imputation of missing values in multivariate normal time series. It accounts for spatial and temporal correlation structures. Temporal patterns can be modelled using an ARIMA(p,d,q). But as with **Amelia**, this functionality is not applicable to univariate data. Providing the imputation function with univariate data throws an error.

Error message from **mtsdi** mniimput function for univariate input:

```
Error in dimnames(S[, , j]) <- list(names[[2]], names[[2]])  :
  'dimnames' applied to non-array
```

**VIM, mice, and imputeR**
Evaluating these packages, it turned out that all three do not accept univariate input. All three give error messages signaling wrong input for non multivariate data. Although for **mice** it seems just a formal issue, since it is capable of using this (simple) univariate imputation techniques: overall mean, linear regression, stochastic regression, random sample, predictive mean matching. To use these functions the time series with the missing values has to be added together with another column without missing data to a data.frame. The input of the second column is arbitrary, since it is only there to bypass the input checking.

**forecast and zoo**
**forecast** and **zoo** were the only packages we found that do both: work for univariate data and provide advanced time series support. Imputation is not the main focus of either package, but since they deal with (univariate) time series, in general they offer these functions as useful additions. The zoo package offers with na.aggregate(), na.StructTS(), na.locf(), na.approx(), na.spline() functions especially for univariate time series. The forecast package has just one but advanced function: na.interp(). Most of the functions tested in our experiments were from these two packages, since these are the only functions for time series data that run with univariate data.

**Custom solutions**
Our focus was to test out of the box imputation solutions in R, that can be applied within one or two lines of code. As it can be seen above, there are only a few functions offering this for univariate time series. Basically, only the functions from **zoo** and **forecast** fit this requirement. But looking not only at out of the box functions, there are additional possibilities for univariate time series imputation in R. But these techniques require a little bit more coding effort.

For example it is also interesting to look at:

- Kalman filter with arima state space model
- Multivariate algorithms on lagged data
- Forecast/Backcast combinations

For our experiments we just used the second option, the multivariate imputation algorithms on lagged data. We did not test more custom code solutions, because our interest was in out of the box functions. But to have an impression, here is an example what the kalman/arima combination would look like.

```
#' Perform imputation with auto.arima and Kalman  filter
  library(forecast)

  ## Airpass used as an  example
  data <- AirPassengers

  ## Set missing values
  data[c(10,13,15)] <- NA

  ## Fit arima model
  fit <- auto.arima(data)

  ## Use Kalman filter
  kal <- KalmanRun(data,  fit$model)
```

```
tmp <- which(fit$model$Z == 1)
id <- ifelse (length(tmp) == 1, tmp[1], tmp[2])

## Fill in the values
id.na <- which(is.na(data))
data[id.na] <- kal$states[id.na, id]
print(data)
```

**Functions used for the experiments**

In this section the R functions used for the experiments are described and it is explained how to apply them in R.

**na.aggregate** (zoo)
'Generic function for replacing each NA with aggregated values. This allows imputing by the overall mean, by monthly means, etc.' (Zeileis and Grothendieck, 2005). In our experiments we just use the overall mean. Computing the overall mean is a very basic imputation method, it is the only tested function that takes no advantage of the time series characteristics. It is very fast, but has clear disadvantages. One disadvantage is that mean imputation reduces variance in the dataset. Also, imputing the overall mean is obviously a bad idea in case of datasets with a strong trend.

```
#' Perform imputation with na.aggregate
  library("zoo")
  na.aggregate(data)
```

**na.locf** (zoo)
'Generic function for replacing each NA with the most recent non-NA prior to it. For each individual, missing values are replaced by the last observed value of that variable' (Zeileis and Grothendieck, 2005). This is probably the most simple algorithm that takes advantage of the time series characteristics of the data. Since there is often a strong relationship between a current observation at point in time $t_n$ and it predecessor at $t_{n-1}$ this can be a quite successful method. Daily temperature data is an example where this works quite well, the temperature for the next day is very likely similar to its predecessor. The method has drawbacks, e.g. when there are huge differences between observation at point in time $t_n$ and it predecessor at $t_{n-1}$ (e.g. time series with strong seasonality).

```
#' Perform imputation with na.locf
  library(zoo)
  # first value can not be imputed with locf
  if ( is.na( data[1] ) ) {
    data[1] <- mean( data, na.rm = T )
  }
  na.locf(data)
```

**na.StructTS** (zoo)
'Generic function for filling NA values using seasonal Kalman filter. The function performs an interpolation with a seasonal Kalman filter. The input time series hereby has to have a frequency' (Zeileis and Grothendieck, 2005).

```
#' Perform imputation with na.StructTS
  library(zoo)
  #The first value of the time series must not be missing
  if ( is.na( data[1 ]) ) {
    data[1] <- mean(data, na.rm = T)
  }
  na.StructTS(data)
```

**na.interp** (forecast)
'Uses linear interpolation for non-seasonal series and a periodic stl decomposition with seasonal series to replace missing values' (Hyndman, 2014). The seasonal component is removed from the time series in the first step, on the remaining component a linear interpolation is done to impute the values and afterward the seasonal component is added again. This methods is especially supposed to be a good fit, where a clear and strong seasonality can be expected.

```
#' Perform imputation with na.interp
  library(forecast)
  na.interp(data)
```

**na.approx** (**zoo**)
'Generic functions for replacing each NA with interpolated values. Missing values (NAs) are replaced by linear interpolation via approx' (Zeileis and Grothendieck, 2005). Since the na.interp function also uses linear interpolation, but first decomposes the time series, it will be interesting to see, if this preprocessing step improves the results compared to just using na.approx.

```
#' Perform imputation with na.approx
  library(zoo)
  na.approx(data, rule = 2)
```

**ar.irmi** (**VIM** + own)
This function is not available from an R package, but can be found in the code belonging to this paper. It is a combination of creating lags for the time series to get a multivariate dataset and afterwards applying the irmi function from the **VIM** package. The function irmi (iterative robust model-based imputation) uses in each step of the iteration one variable as a response variable and the remaining variables serve as the regressors. (Templ et al., 2013)

```
#' Perform imputation an lagged data with  ar.irmi
  library(VIM)
  ar.irmi(data, lags = 10)
```

The ar.irmi function internally first calls the create.lags function in order to create the lags and then it calls the irmi function. For creating lags we did not use the function lag from **stats**, because instead of computing a lagged version of the time series, it's shifting back the time base by a given number of observations. The lag function from **zoo** also did not fit our needs, so that we created an own function. Although we did not use it, we want to mention the **DataCombine** (Gandrud, 2015) package, whose slide function would have perfectly fitted our needs. We just discovered the package too late and had already finished our experiments.

```
#' Transform a univariate time series to a matrix with lags as  columns.
#' @param data The time series.
#' @param lags The maximum amount of lags to be  created.
#' @return A data frame with the lags in the  columns.

create.lags <- function(data, lags = 0) {
    if (lags < 1) {
        warning("No lags  introduced.")
        return(data)
    }
    data.new <- data.frame(data[(lags+1):length(data)])
    cnames <- "x"
    for (i.lag in 1:lags) {
        ind <- (lags + 1 - i.lag) : (length(data) - i.lag)
        data.new <- cbind(data.new, data[ind])
        cnames <- c(cnames, paste("lag", as.character(i.lag), sep = "_"))
    }
    colnames(data.new) <- cnames
    return(data.new)
}
```

# Experiments

## Description

To compare the different R functions, we performed experiments with different time series datasets. All code from our experiments can be found online under http://github.com/SpotSeven/uniTSimputation.

In order to quantify the performance of imputation algorithms the following steps have to be done:

1. Take a complete time-series $ts_{compl}$
2. Create missing values in $ts_{compl}$ to get $ts_{incompl}$
3. Apply the algorithms to $ts_{incompl}$ to get $ts_{imp}$
4. Compare the differences between $ts_{compl}$ and $ts_{incompl}$

In our paper we examine the special case of imputation of univariate time series and compare several state of the art solutions available within R packages. To be able to compare results we took a two step approach. In the first step we artificially deleted values in complete datasets with a missing data simulation function we developed based on the occurrence of missing data in a real life example (MCAR). The second step was to apply the earlier introduced imputation functions and to evaluate the performance of the algorithms in terms of Root Mean Square Error (RMSE) and Mean Absolute Percentage Error (MAPE) for the imputed values.

The chosen datasets for the experiments are described in section Univariate Time Series. For step 2 a defined function that creates missing values with a realistic distribution is needed. Details of our missing value creation function can be found in section Missing Data. All algorithms used in step 3 are described in detail in section Univariate time series imputation. For step 4, we need error metrics to evaluate which algorithm gives the best result. The metrics we used are describes in the following subsection.

The experiments were performed with four different missing data rates (0.1, 0.3, 0.5, 0.7). The rate represents the $\lambda$ value of the exponential distribution (which is the foundation of the missing data creation function). The same rate can hereby lead to a slightly differing absolute value of observations missing. Since the performance of the algorithms also depends on which specific observations are deleted by the missing data creation function, we ran this function with 25 different random seeds. So all in all, our experiments were performed with 25 (different random seeds) * 4 (different rates) * 6 (different algorithms) = 600 runs for each dataset.

## Imputation Targets and Measurements

As error metric we are relying on MAPE and RMSE, two very common metrics. Adding the MAPE as error metric is important because for datasets with a strong trend, the RMSE may not be the best metric. Of course, it depends on the requirements of the specific application (as specified by a user) whether RMSE or MAPE represent the quality of the algorithm well. Hence, we added both metrics.

For example, the air passengers datasets starts with very low values and ends up with very high values. The missing observations later in the dataset would have an very high impact on the RMSE, while the first ones would only have a small impact. In such cases a error metric based on the difference in percent between imputed value and real value can give a more realistic view than a metric that calculates the difference as absolute value.

**RMSE** The *root mean square error* (RMSE) between the imputed value $\bar{y}$ and the respective true value time series $y$, i.e.,

$$\text{RMSE}(\bar{y}, y) := \sqrt{\frac{\sum_{t=1}^{n} (\bar{y}_t - y_t)^2}{n}}$$

**MAPE** The *mean absolute percentage error* (MAPE) between the imputed value $\bar{y}$ and the respective true value time series $y$, i.e.,

$$\text{MAPE}(\bar{y}, y) := \frac{\sum_{t=1}^{n} \frac{|\bar{y}_t - y_t|}{|y_t|}}{n}$$

## Results of the experiments

For each dataset a MAPE and a RMSE figure comparing different R functions for univariate time series imputation will be shown. One point in the figure is equivalent to one imputation result (given as MAPE or RMSE) for one variation of the time series (same series, but with differing missing values, due to different random seeds). The colors in the figures mark different missingness rates.

### Airpass

Looking at figure 4 it can be seen, that na.StructTS and na.interp show the best results. This is probably because they can handle seasonality in the data better than the other algorithms. Airpass has both seasonality and trend. The trend is probably the reason why mean imputation (na.aggregate) shows the poorest results (huge differences in the mean level). The other three algorithms are located in the middle between this two poles. As can be seen in figure 5, RMSE and MAPE lead to similar results.

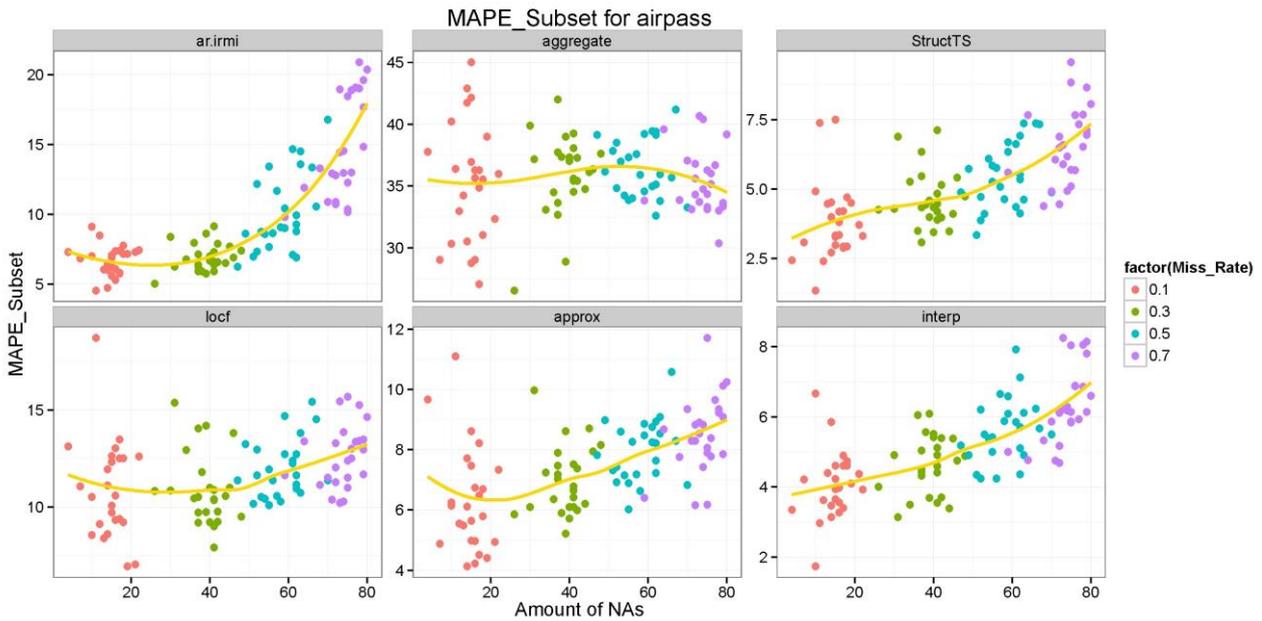

**Figure 4:** MAPE imputation results for airpass series.

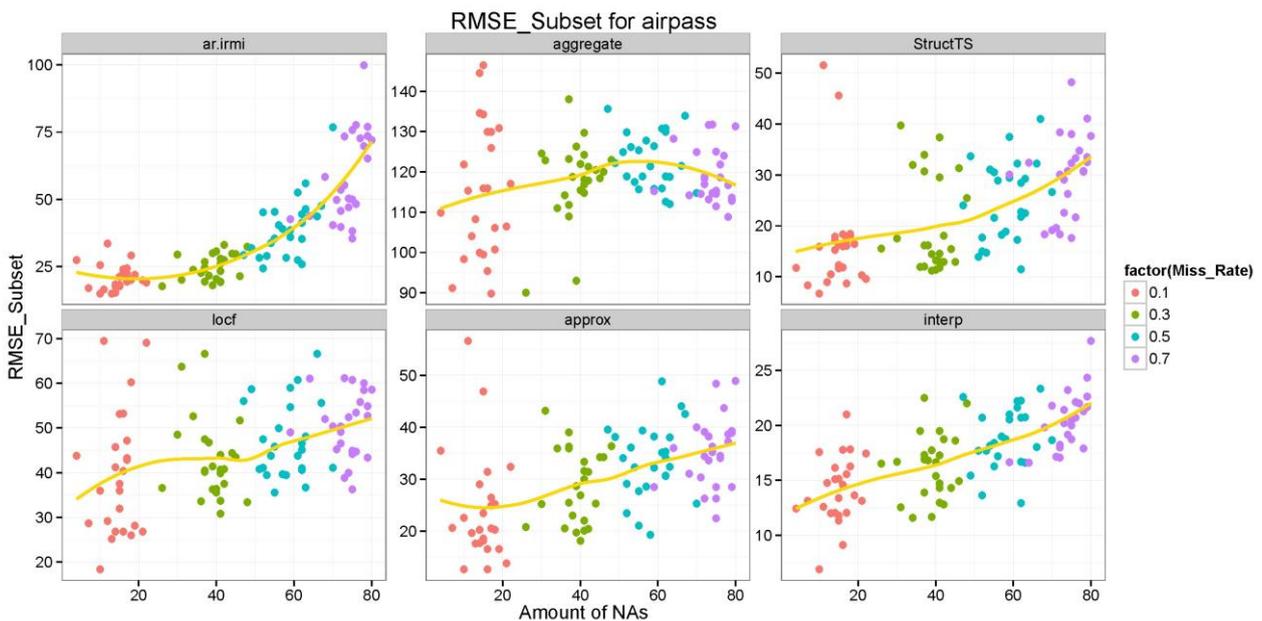

**Figure 5:** RMSE imputation results for airpass series.

**beersales**

The beersales series also has a clear seasonality. Thus, it is not surprising, that na.interp and StructTS again show the best results. Here, the results of the na.interp function seem even a little bit better than the ones from na.StructTS. StructTS shows more negative outliers, while na.interp gives more constant results. Approx and ar.irmi follow this two algorithms quite close, while locf and na.aggregate come in last with some distance. Since this series has no clear trend, na.aggregate is not as far behind the other algorithms as for the airpass dataset. It is also interesting to compare na.approx and na.interp, both functions use linear interpolation, but na.interp does this on seasonal loess decomposed data. This seems to give a small gain. As can be seen by looking at the two figures also for this dataset, RMSE and MAPE lead to the same results.

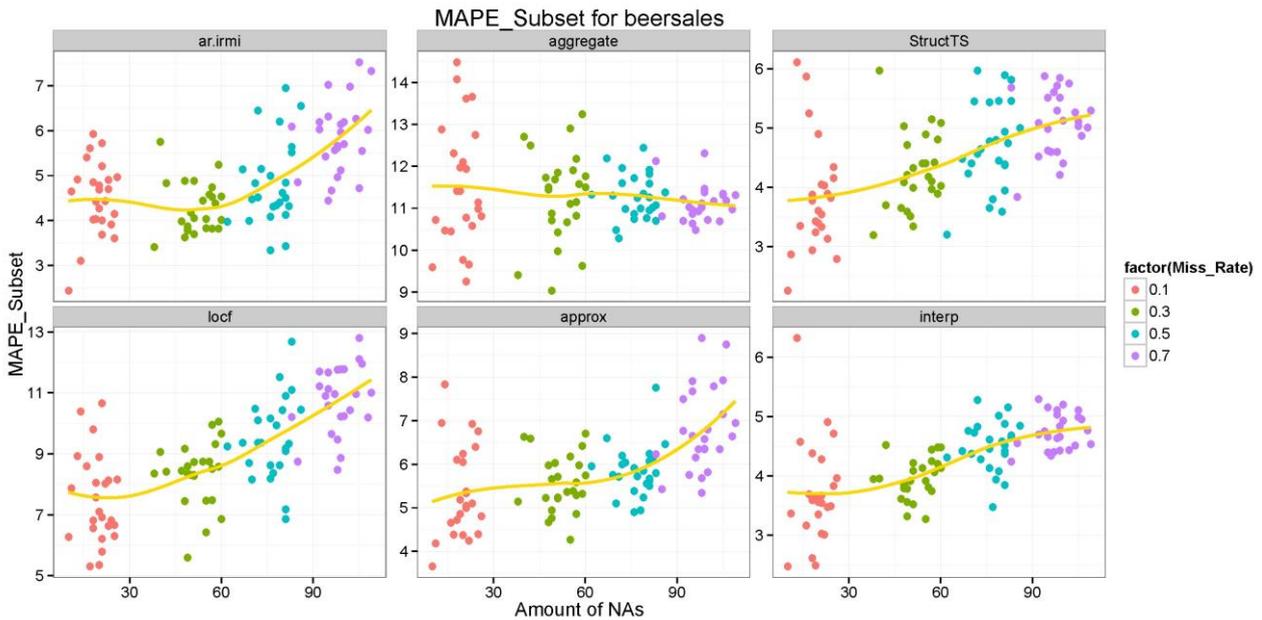

**Figure 6:** MAPE imputation results for beersales series.

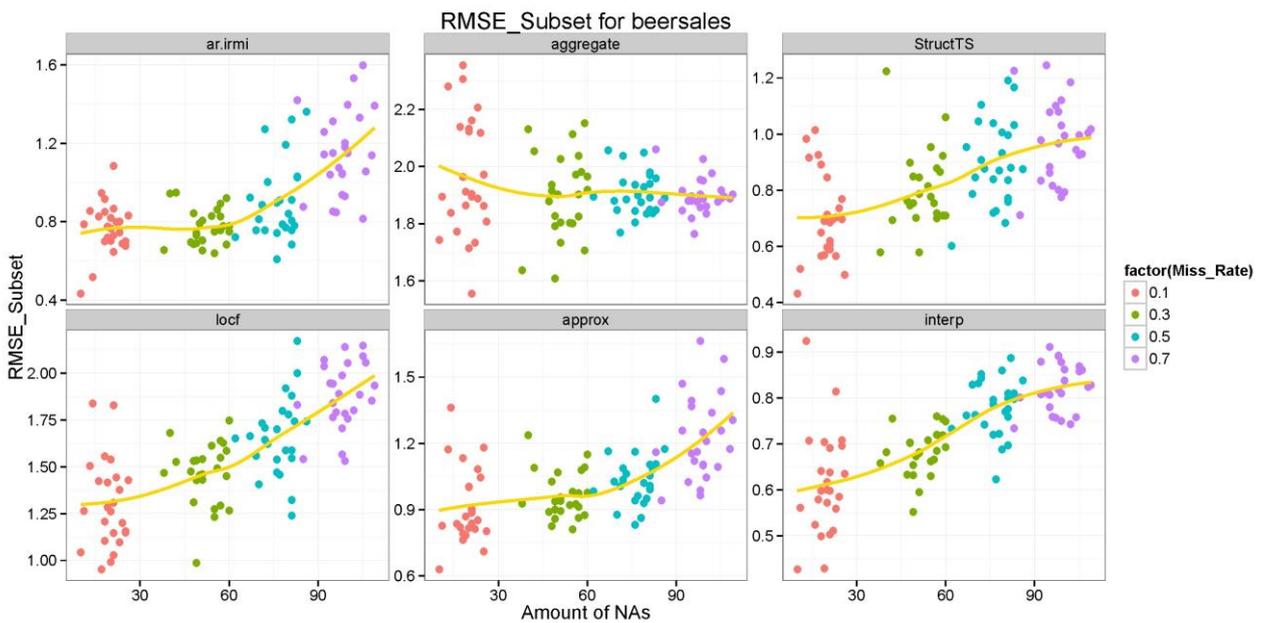

**Figure 7:** RMSE imputation results for beersales series.

**google**
The google dataset was the series without seasonality and without trend, being nearly white noise. All algorithms show bad results on this dataset, as it could be expected. It is remarkable that na.StructTS and ar.irmi show a few very high outliers. The na.approx function gives here exactly the same results as na.interp, since there is no seasonality, which na.interp could work on. Because of the outliers in na.StructTS and ar.irmi, these two are the worst algorithms on this dataset. The best algorithms seem to be both interpolation algorithms and na.locf. But the na.aggregate function seems to be nearly on their level.

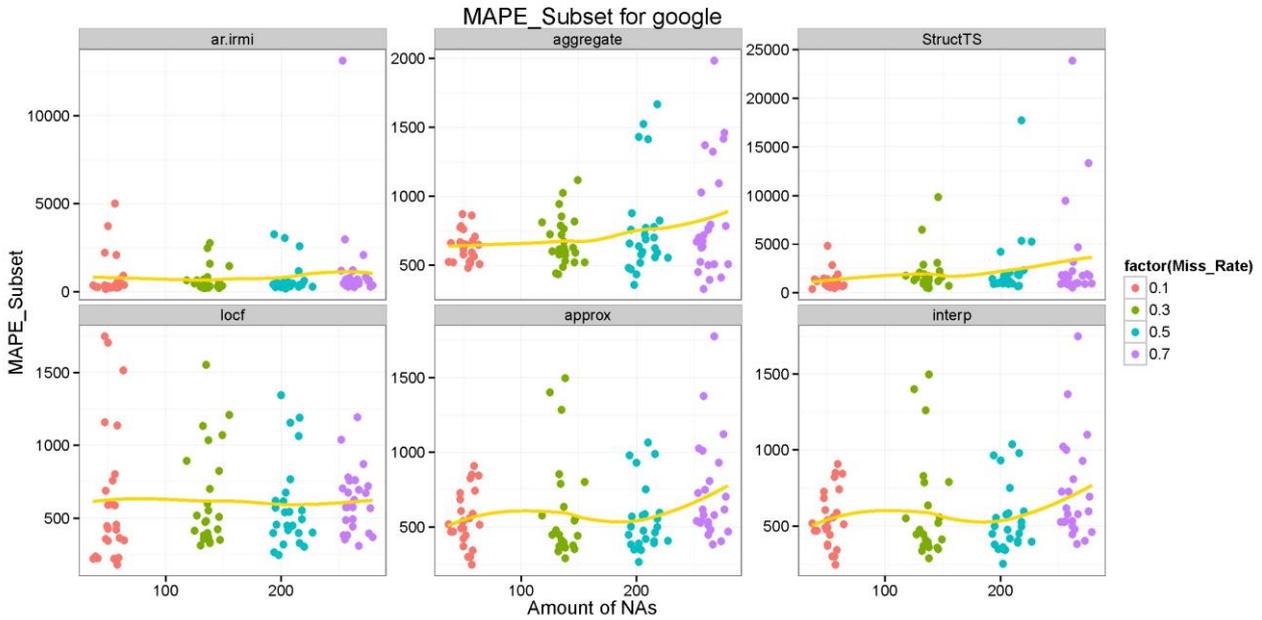

**Figure 8:** MAPE imputation results for google series.

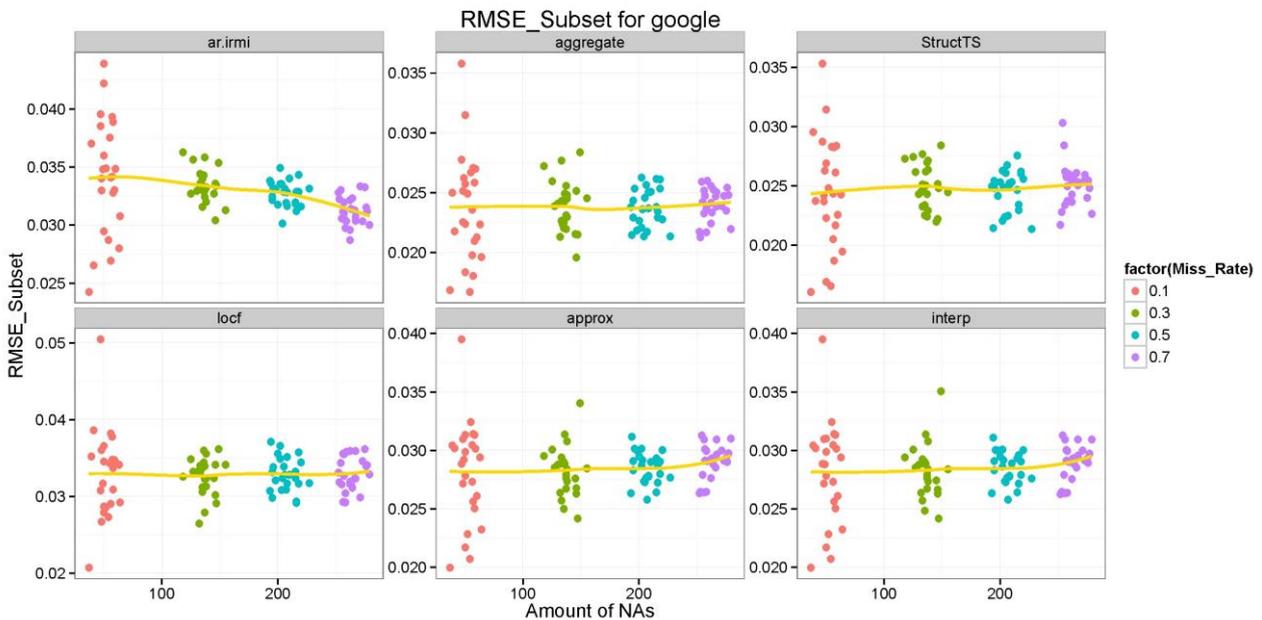

**Figure 9:** RMSE imputation results for google series.

**SP**
The SP dataset was the series with just trend and without seasonality. The best algorithm on this dataset seems to be na.approx. This is a quite interesting result, because this means that the seasonal adaption in na.interp makes the results worse. Also StructTS shows poorer results than a normal interpolation. The mean is again the worst algorithm to choose for this dataset. Last observation carried forward lies far away from the results of na.aggregate but still not even close to the best algorithms.

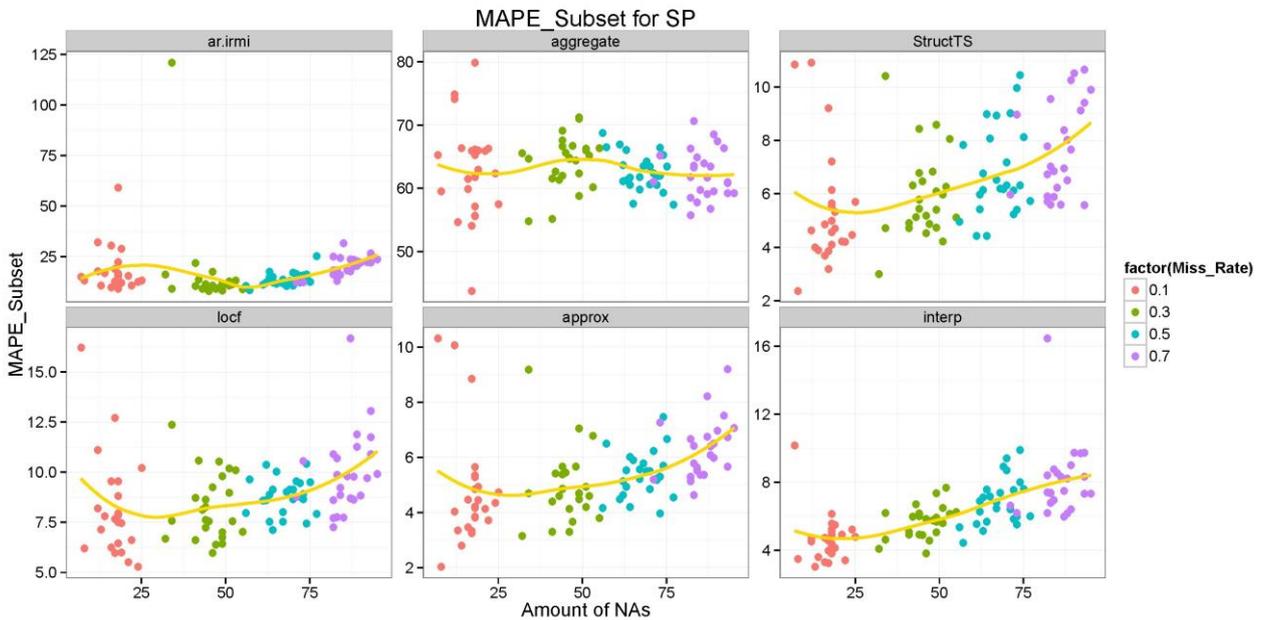

**Figure 10:** MAPE imputation results for SP series.

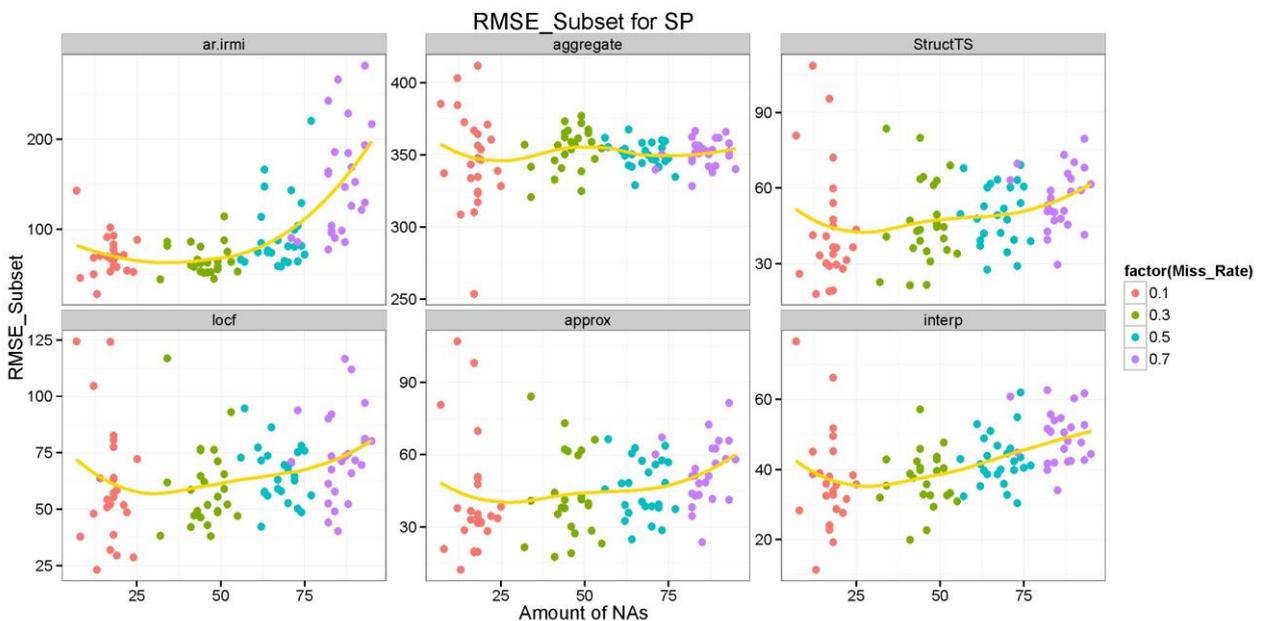

**Figure 11:** RMSE imputation results for SP series.

**Running Times**
The running times are depicted in figure 12, na.aggregate is clearly the fastest algorithm. The ar.irmi is the solution that needs the most computation time. Keeping in mind that ar.irmi showed only mediocre results, the computation time seems not to be worth it. What is very interesting is that StructTS is several times slower than na.interp. Both showed nearly identical results for imputation, so na.interp seems overall the better choice. The na.interp function is also not that much slower than na.approx, so the loess decomposition seems not to be very demanding in terms of computing time.

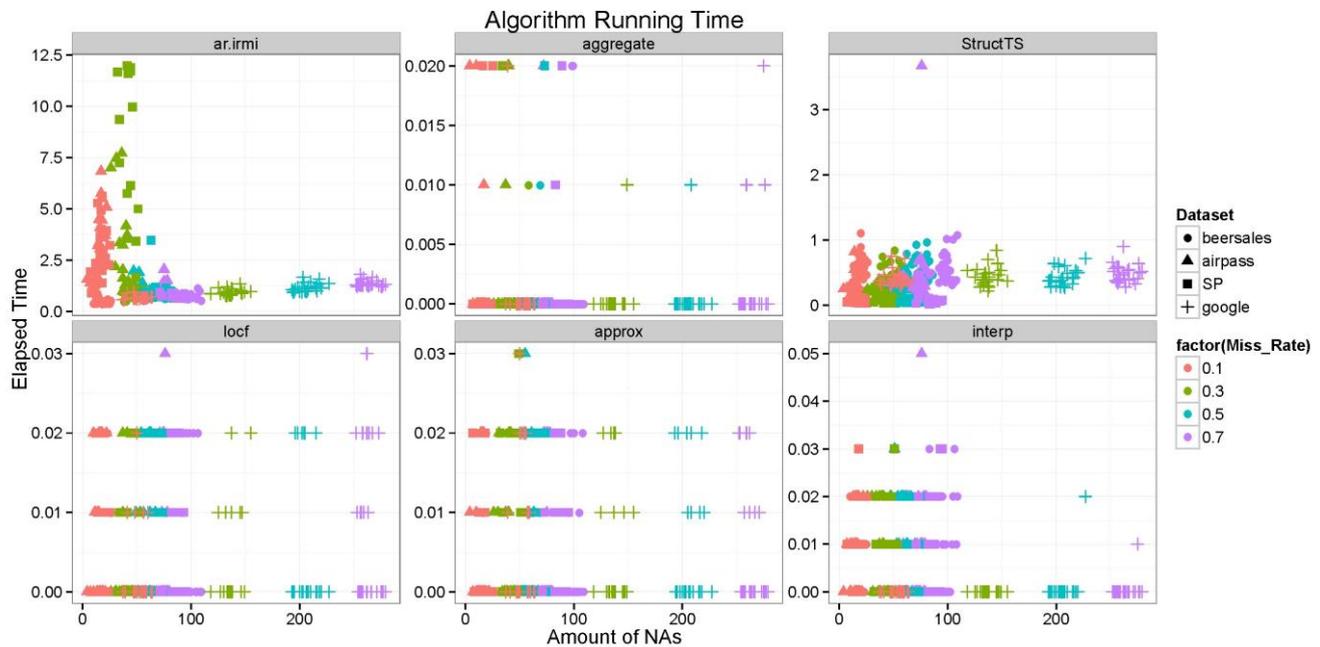

**Figure 12:** Running time of the algorithms in seconds.

## Summary

As shown in sections Univariate Time Series and Missing Data, handling univariate time series with standard imputation algorithms may not be the best solution or even impossible. Univariate time series require special treatment, since their characteristics are different from multivariate, non-time series datasets. Instead of covariates, time dependencies have to be employed to perform an effective imputation.

There are basically three different approaches to perform imputation for univariate time series (see also section Univariate time series imputation): Using univariate imputation algorithms without special treatment of time series characteristics, explicitly using time series algorithms or using multivariate algorithms on lagged data.

In our experiments the algorithms especially tailored to (univariate) time series imputation performed for each tested dataset better or at least equal to all other algorithms. The results for the univariate (non-time series) imputation algorithms were clearly inferior to the results of the other algorithms. The results for the multivariate algorithm using lagged data was, compared over all datasets, somewhere in the middle between the other two algorithm categories. It would be interesting to do additional tests with more algorithms from this category.

In R, there are currently not too many methods that provide imputation for univariate time series out of the box. In the future, having an R package that unites several tools and algorithms specifically for univariate imputation would be an improvement. From the R implementations tested in this paper, **na.interp** from the **forecast** package and **na.StructTS** from the **zoo** package showed the best overall results. The simple methods **na.locf** and **na.aggregate** (also from package **zoo**) were the leaders in terms of computation time. But, depending on the dataset, they produced very misleading results and should be applied with caution. The approach of using a multivariate algorithm (**irmi** from package **VIM**) on lagged series showed mediocre results, and was extremely slow compared to all other solution.

*Steffen Moritz*
*Cologne University of Applied Sciences*
*Cologne, Germany*
steffen.moritz10@gmail.com

*Alexis Sardá*
*Cologne University of Applied Sciences*
*Cologne, Germany*
alexis.sarda@gmail.com

*Thomas Bartz-Beielstein*
*Cologne University of Applied Sciences*
*Cologne, Germany*
bartz.beielstein@fh-koeln.de

*Martin Zaefferer*
*Cologne University of Applied Sciences*

*Cologne, Germany*
martin.zaefferer@fh-koeln.de

*Jörg Stork*
*Cologne University of Applied Sciences*
*Cologne, Germany*
joerg.stork@fh-koeln.de